\documentclass[times, 10pt,twocolumn]{article}
\usepackage{latex8}
\usepackage{times}
\usepackage{algorithm}
\usepackage{algorithmic}
\usepackage{graphicx}
\usepackage{graphics} 
\usepackage{epsfig} 

\begin{document}

\title{Statistical Cryptography using a Fisher-Schr\"{o}dinger Model}

\author{R. C. Venkatesan\\
Systems Research Corporation\\ Aundh, Pune 411007, India \\
ravi@systemsresearchcorp.com\\}

\maketitle
\thispagestyle{empty}

\begin{abstract}
   A principled procedure to infer a hierarchy of statistical
distributions possessing ill-conditioned eigenstructures, from
incomplete constraints, is presented.  The inference process of the
\textit{pdf}'s employs the Fisher information as the measure of
uncertainty, and, utilizes a semi-supervised learning paradigm based
on a measurement-response model. The principle underlying the
learning paradigm involves providing a quantum mechanical
connotation to statistical processes. The inferred \textit{pdf}'s
constitute a statistical host that facilitates the
encryption/decryption of covert information (code). A systematic
strategy to encrypt/decrypt code via unitary projections into the
\textit{null spaces} of the ill-conditioned eigenstructures, is
presented. Numerical simulations exemplify the efficacy of the
model.
\end{abstract}

\Section{Introduction}

This paper accomplishes a two-fold objective.  First, a systematic
methodology to infer from incomplete constraints, a hierarchy of
statistical distributions corresponding to the \textit{multiple
energy states} of a time independent Schr\"{o}dinger-like equation
(TIS-lE), is presented. By definition, the case of incomplete
constraints corresponds to scenarios where the number of constraints
(physical observables) is less than the dimension of the
distribution.  The inference procedure employs a semi-supervised
learning paradigm, based on a measurement-response model that
utilizes the Fisher information (FI) as the measure of uncertainty.

The time independent Schr\"{o}dinger equation (TISE) is a
fundamental equation of physics, that describes the behavior of a
particle in the presence of an external potential [1]
\begin{equation}\label{1}
\underbrace { - \frac{{\hbar ^2 }}{{2m}}\frac{{d^2 \psi \left( x
\right)}}{{dx^2 }} + V\left( x \right)\psi \left( x \right)}_{H^{QM}
\psi \left( x \right)} = E\psi \left( x \right). \
\end{equation}
Here, $ \psi \left( x \right) $ is the wave function, $ E $ is the
total energy eigenvalue,$ V(x) $ is the external potential, and, $
H^{QM} $ is the TISE Hamiltonian. The constants $ \hbar $ and $ m $
are the Planck constant and the particle mass, respectively.  In
time independent scenarios, Lagrangians containing the FI as the
measure of uncertainty, yield on variational extremization an
equation similar to the TISE, i.e. the TIS-lE [2] \footnote{This
property is the \textit{raison d'etr\'{e}} for the phrase
"Fisher-Schr\"{o}dinger model"}. \textit{The TIS-lE provides a
quantum mechanical connotation to a statistical process}.

Next, a self-consistent strategy to project covert information into
the \textit{null spaces} of ill-conditioned eigenstructures
possessed by the inferred host statistical distributions
corresponding to the \textit{multiple energy states} of the TIS-lE,
is described. The strategy of unitary projection of covert
information into the \textit{null spaces} of the ill conditioned
eigenstructures of a hierarchy of statistical distributions, has
been recently studied for host probability density functions
(\textit{pdf}'s, hereafter) inferred using the maximum entropy
(MaxEnt) principle [3].

The selective projection of covert information into a hierarchy of
statistical distributions implies that the dimension of the covert
information is greater than that of any single host distribution.
This selective projection endows the code\footnote{The terms covert
information and code are used interchangeably.} with multiple layers
of security, without altering the host statistical distributions.
The present paper accomplishes the task of achieving both symmetric
and asymmetric cryptography [4] via a judicious amalgamation of
statistical inference using an information theoretic semi-supervised
learning paradigm, quantum mechanics, and, the theory of unitary
projections.

In summary, the semi-supervised learning paradigm is utilized to
infer the statistical hosts possessing ill-conditioned
eigenstructures.  The code is then projected into the \textit{null
spaces} of these ill-conditioned eigenstructures.  Another example
of the use of learning theory in cryptosystems, albeit within a
different context, is described in [5].

\subsection{The TIS-lE}

Consider a measured random variable $ y = \left( {y_1 ,...,y_N }
\right) $, parameterized by $ \theta  = \left( {\theta _1
,...,\theta _N } \right) $ (the "true" value). A
\textit{fluctuation}, i.e. a random variable $ x = \left( {x_1
,...,x_N } \right) $, defined by $ x = y - \theta $ is introduced.
For translational (or shift) invariant families of distributions, $
p\left( {y\left| \theta \right.} \right) = p\left( {y - \theta }
\right) = p\left( x \right) $.  The particular form of the FI that
is chosen is the trace of the FI matrix for independent and
identically distributed (\textit{iid}, hereafter) data \footnote{The
FI matrix for \textit{iid} data has vanishing off-diagonal elements
(e.g., Appendix B in [2])}. This is referred to as the Fisher
channel capacity (FCC) [2]. The FCC is: $ I^{FCC} = \int {dyp\left(
{\left. y \right|\theta } \right)} \left( {\frac{{\partial \ln
p\left( {\left. y \right|\theta } \right)}}{{\partial \theta }}}
\right)^2 = \int {dxp\left( x \right)} \left( {\frac{{\partial \ln
p\left( x \right)}}{{\partial x}}} \right)^2 $  under translational
invariance [2,6]. The probability amplitude (wave function) relates
to the \textit{pdf} as $ \psi \left( x \right) = \sqrt {p\left( x
\right)} $.  The FCC acquires the compact form $ I^{FCC} =
4\sum\limits_{n = 1}^N {\int {dx_n \left( {\frac{{\partial \psi
\left( {x_n } \right)}}{{\partial x_n }}} \right)} ^2 }= 4\int
{dx\left( {\frac{{\partial \psi \left( x \right)}}{{\partial x}}}
\right)} ^2 $.  The form of the FCC is essential to the formulation
of a variational principle.  Within the framework of a
measurement-response model, this implies that the observer who
initiates the measurements, collects the response in the form of
\textit{iid} data. In many practical scenarios, the response of a
system to measurements is not obliged to be \textit{iid}.  The
presence of correlations contribute to off-diagonal elements in the
FI matrix formed by the observer. These correlations may be
mitigated, thereby eliminating the off-diagonal elements of the FI
matrix, by performing ICA (or an equivalent procedure) as a
pre-processing stage.

Consider a Lagrangian of the form
\begin{equation}
\begin{array}{l}
 L^{FCC}  = 4\int {dx\left( {\frac{{\partial \psi \left( x \right)}}{{\partial x}}} \right)} ^2  +  \\
  + \underbrace {\int {dx\sum\limits_{i = 1}^M {\lambda _i \Theta _i \left( x \right)\psi ^2 \left( x \right)}  - \lambda _o \int {dx\psi ^2 \left( x \right)} } }_{ - J\left[ x \right]}, \\
 \end{array}
 \end{equation}
$ M < N $ (incomplete constraints), where the Lagrange multiplier
(LM) $ \lambda _o $ corresponds to the probability density function
(\textit{pdf}, hereafter) normalization condition $ \int {dx\psi ^2
\left( x \right)}  = 1 $. The LM's $ \lambda _i ;i = 1,...,M $
correspond to actual (physical) constraints of the form $ \int
{dx\Theta _i \left( x \right)\psi ^2 \left( x \right)} =
\left\langle {\Theta _i \left( x \right)} \right\rangle  = d_i $.
Here, $ \Theta _i \left( x \right) $ are operators, and, $ d_i $ are
the constraints (physical observables). This work considers
constraints of the geometric moment type: $ \Theta _i \left( x
\right) = x^i ;i = 0,...,M $. Here, (2) resembles the usual MaxEnt
Lagrangian with the FCC replacing the Shannon entropy. In (2), the
FCC is ascribed the role akin to the kinetic energy. The constraint
terms manifest the potential energy. Variational extremization of
(2) yields the minimum Fisher information (MFI) principle [7]
\begin{equation}\label{3}
\underbrace { - \frac{{d^2 \psi \left( x \right)}}{{dx^2 }} +
\frac{1}{4}\sum\limits_{i = 1}^M {\lambda _i \Theta _i \left( x
\right)} \psi \left( x \right)}_{H^{FI} \psi \left( x \right)} =
\frac{{\lambda _o }}{4}\psi \left( x \right),\
\end{equation}
where $ H^{FI} $  is an empirical Hamiltonian operator. Here, (3) is
referred to as a TIS-lE.   Note that the probability amplitudes are
taken as being real quantities.  This assumption is tenable since
the model is spatially one dimensional in the continuum, and is time
independent. Comparing the TIS-lE with the TISE immediately reveals
that the constraint terms $ V\left( x \right) =
\frac{1}{4}\sum\limits_{i = 1}^M {\lambda _i } \Theta _i \left( x
\right) = \sum\limits_{i = 1}^M {\tilde \lambda _i } x^i $,
constitute an empirical pseudo-potential. The normalization LM and
the total energy eigenvalue relate as $ \lambda _o  = 4E $. Further,
the constants in the TISE relate as $ {{\hbar ^2 } \mathord{\left/
 {\vphantom {{\hbar ^2 } {2m}}} \right.
 \kern-\nulldelimiterspace} {2m}} = 1 $.

Solution of the TISE as an eigenvalue problem yields a number of
\textit{energy states} characterized by distinct values of $ E $.
These comprise the equilibrium state (\textit{zero-energy state})
characterized by a Maxwellian distribution, and, higher energy
\textit{excited states} (non-equilibrium states). The wave functions
are a superposition of Hermite-Gauss solutions. \textit{By virtue of
its similarity to the TISE, the TIS-lE ``inherits" these
\textit{energy states }within an information theoretic context}.
This feature permits the projection of covert information into
\textit{multiple energy states} of the TIS-lE, for an empirical
pseudo-potential that approximates a TISE physical potential.

Employing the TIS-lE to infer \textit{pdf}'s from incomplete
constraints, requires an accurate evaluation of the LM's.  This is
accomplished in this paper through a semi-supervised learning
paradigm, that iteratively couples the solution of (3) with the
minimization of a Lagrangian that manifests a measurement-response
model.

This procedure represents, in certain aspects, an extension of the
optimization procedure employed to achieve \textit{quantum
clustering} using the TISE [8]. In the case of \textit{quantum
clustering}, the TISE probability amplitude/wave function is
approximated by a non-parametric estimator (Parzen windows), and,
the potential $ V(x) $ is determined via a steepest descent in
Hilbert space. In contrast, the semi-supervised paradigm presented
in this paper achieves reconstruction of \textit{pdf}'s (the inverse
problem of statistics) without any \textit{a-priori} strictures
placed on the probability amplitudes of (3).  The
Fisher-Schr\"{o}dinger model has been employed within a statistical
setting in a number of studies ranging from quantum statistics to
fuzzy clustering [9]. Within the context of securing covert
information, the above features endow the statistical
enryption/decryption strategy with a fundamental physical
connotation.

\subsection{The Dirac notation}

This paper utilizes the Dirac \textit{bra-ket} notation [10] to
describe linear algebraic operations in a compact form.  By
definition, a \textit{ket} $ \left|  \bullet  \right\rangle $
denotes a column vector, and, a \textit{bra} $ \left\langle  \bullet
\right| $ denotes a row vector.  The scalar inner product and the
projection operators are described by $ \left\langle { \bullet }
 \mathrel{\left | {\vphantom { \bullet   \bullet }}
 \right. \kern-\nulldelimiterspace}
 { \bullet } \right\rangle $, and the outer product $
\left|  \bullet  \right\rangle \left\langle  \bullet  \right| $,
respectively.  The expectation evaluated at the $ \varepsilon^{th} $
\textit{energy state} is $ \left\langle  \bullet  \right\rangle
_\varepsilon $.

 \section{Semi-supervised Learning Paradigm}

\subsection{Theory}
 The task of density estimation involves the iterative determination of the LM's and probability amplitudes of the TIS-lE (3).  In the MaxEnt and MFI theories, the observer is external to the system.
 The present work reconstructs the host \textit{pdf}'s using a semi-supervised learning paradigm, by incorporating a
\textit{participatory observer}. This is accomplished by positioning
the \textit{participatory observer} in a \textit{measurement space}
characterized by the amplitude $ \psi^\varepsilon (x) $, performing
unbiased measurements [2, 11] on a given physical system (data).

The \textit{system space}, inhabited by the physical system subject
to measurements, is characterized by an amplitude $ \phi^\varepsilon
(\tilde x) $.
 Here, $ x $ and $ \tilde x $ are the conjugate basis
coordinates of the \textit{measurement space} and \textit{system
space}, respectively.  Herein, the mutually conjugate spaces are
taken to be the Cartesian coordinate and the linear momentum.
Setting $ \tilde x=\mu $, the commutation relation is $ \left[
{x,\mu } \right] = i\hbar $, respectively [1]. The group for the
basis change is $ G = - i\hbar \frac{d}{{dx}} $. The corresponding
Hermitian unitary operator is $ U = e^{ - ai\hbar \frac{d}{{dx}}} $,
where $ a $ is the group parameter of infinitesimal transformations.
Within the present scenario, $ U\psi^\varepsilon \left[ x \right] =
\varphi^\varepsilon \left[ {\mu} \right] $.  On the basis of the
above discussions, it is easily proven that $ I_\varepsilon^{FCC}
\left[ x \right] = 4\int {dx\left( {\frac{{d\psi \left( x
\right)}}{{dx}}} \right)} ^2  = \frac{4}{{\hbar ^2 }}\left\langle
{\mu ^2 } \right\rangle_\varepsilon  = 4\left\langle {\frac{{\mu ^2
}}{{2m}}} \right\rangle_\varepsilon  = I_\varepsilon^{FCC} \left[
\mu \right];{\raise0.7ex\hbox{${\hbar ^2 }$} \!\mathord{\left/
 {\vphantom {{\hbar ^2 } {2m}}}\right.\kern-\nulldelimiterspace}
\!\lower0.7ex\hbox{${2m}$}} = 1 $.  In the non-relativistic limit,
the kinetic energy of a particle is $ T = \frac{{\mu ^2 }}{{2m}} $
[1]. For TIS-lE polynomial pseudo-potentials of the form $ V\left( x
\right) = \sum\limits_{i = 1}^M {\lambda _i^\varepsilon x^i } $, the
quantum mechanical virial theorem [9, 12, 13] yields
\begin{equation}\label{3}
I_\varepsilon ^{FCC} \
 = 2\left\langle
{x\frac{{dV\left( x \right)}}{{dx}}} \right\rangle _\varepsilon   =
2\sum\limits_{i = 1}^M {i\lambda _i^\varepsilon d_i^\varepsilon } \
\end{equation}

\textit{The unitary relation between the amplitudes in conjugate
spaces results in the potential energy term in (2), $
J_\varepsilon\left[ x \right] $, being manifested as an empirical
representation of the FCC}. Specifically, $ I_\varepsilon ^{FCC}
\left[ x \right] = J_\varepsilon  \left[ x \right] $.  Each
measurement (or set of measurements) initiated by the observer at a
specific juncture, perturbs the amplitude of the \textit{system
space} as $ \delta \varphi ^\varepsilon \left( {\mu} \right) $. For
mutually conjugate spaces related by a unitary transform, this
results in a perturbation $ \delta \psi ^\varepsilon \left( x
\right) = \delta \varphi ^\varepsilon \left( {\mu} \right) $ of the
\textit{measurement space}. It is at this juncture that the observer
constructs the FCC for \textit{iid} data.   Consequently, $ \delta
I_\varepsilon ^{FCC} \left[ x \right] = \delta J_\varepsilon \left[
x \right] $ [2]. Such models are known as measurement-response
models [14].

Incorporation of a \textit{participatory observer} results in a
\textit{zero-sum game} [15] of information acquisition played
between the observer and the system under observation. The observer
seeks to maximize her/his information about the system.
Simultaneously, the \textit{system space} is inhabited by a
\textit{demon}, remniscent to the Maxwell \textit{demon}, who seeks
to minimize this information transfer. This \textit{zero-sum game}
between the observer and the \textit{demon} is hereafter referred to
as the \textit{Fisher game}.

Game theoretic studies in MaxEnt and MFI follow the traditional
pattern of having the \textit{arbiter}, who assigns strategies to
the players, residing external to the system. \textit{In this paper,
the probe measurements initiated by the observer constitute the
arbiter, and, the probability amplitudes manifest the strategies.} A
future publication treats the game theoretic aspects of the
semi-supervised learning paradigm, within the ambit of the
\textit{bounded rationality theory} [16].

The incomplete constraints $ {d_i^\varepsilon },i=1,...,M $ are
evaluated as the moments of the Cartesian coordinates at each
\textit{energy state} $ \varepsilon $, by solving the TISE (1) as an
eigenvalue problem on a lattice, for an \textit{a-priori} specified
physical potential. \textit{The incomplete constraints represent the
only manifestation of the target values of the
amplitudes/\textit{pdf}'s, made available to the designer at the
commencement of the inference procedure}.

The host \textit{pdf }inference is solved by an iterative
optimization process \textit{In this paper, the host pdf is
independently inferred for each energy level $ \varepsilon $}.  The
optimization process couples the solution of the TIS-lE (3), with
the steepest descent minimization of an empirical quantity known as
the \textit{residue}. The \textit{residue} represents the
discrepancy between the value of $ J_\varepsilon  \left[ x \right] $
in (2) evaluated at an intermediate \textit{iteration level} $ l $
for a specific \textit{energy state}, and, the value of the exact
(target) FCC expressed at the same \textit{iteration level} $ l $.

In (4), the LM's $ \lambda _i^\varepsilon $ are target values of the
optimization process, which are unknown at the commencement of the
inference procedure. Here, (4) is made consistent with the iteration
process by specifying the relation between the target values of the
LM's, and, the LM's at some intermediate iteration level $ l $ as
\begin{equation}\label{3}
\begin{array}{l}
2 \sum\limits_{i = 1}^M {i\lambda _i^\varepsilon  }  \cong
\sum\limits_{i = 1}^M {\lambda _{i,l}^\varepsilon  \left\langle
{{\psi _l^\varepsilon  }}
 \mathrel{\left | {\vphantom {{\psi _l^\varepsilon  } {\psi _l^\varepsilon  }}}
 \right. \kern-\nulldelimiterspace}
 {{\psi _l^\varepsilon  }} \right\rangle }_\varepsilon \
\end{array}
\end{equation}

\textit{Here, (5) is critical to the optimization process since it
infuses a representation of the target response state into the
iterative procedure}.  The final values of the expectation of the
amplitudes satisfy $ \left\langle {{\psi _{l = final}^\varepsilon
}}
 \mathrel{\left | {\vphantom {{\psi _{l = final}^\varepsilon  } {\psi _{l = final}^\varepsilon  }}}
 \right. \kern-\nulldelimiterspace}
 {{\psi _{l = final}^\varepsilon  }} \right\rangle _\varepsilon   = 1 $.  Note that the expectation $ {\left\langle {{\psi _l^\varepsilon  }}
 \mathrel{\left | {\vphantom {{\psi _l^\varepsilon  } {\psi _l^\varepsilon  }}}
 \right. \kern-\nulldelimiterspace}
 {{\psi _l^\varepsilon  }} \right\rangle }_\varepsilon
$ is not assumed to be unity. Combining (4) and (5) allows the
target value of the FCC to be manifested at some intermediate
iteration level $ l $. At the $ \ l^{th} \ $ iteration level, the
term $ J_\varepsilon[x] $ in (2) is
\begin{equation}
\begin{array}{l}
 J \left( {\lambda _{i,l}^\varepsilon  } \right) = \lambda _{o,l}^\varepsilon  \left\langle {{\psi _l^\varepsilon  }}
 \mathrel{\left | {\vphantom {{\psi _l^\varepsilon  } {\psi _l^\varepsilon  }}}
 \right. \kern-\nulldelimiterspace}
 {{\psi _l^\varepsilon  }} \right\rangle_\varepsilon  - \sum\limits_{i = 1}^M {\lambda _{i,l}^\varepsilon  \left\langle {\psi _l^\varepsilon  } \right|\Theta \left( x \right)\left| {\psi _l^\varepsilon  } \right\rangle }_\varepsilon  \\
 \end{array}
\end{equation}

The \textit{residue} at the $ l^{th} $ iteration level, re-scaled
with respect to $ \left\langle {{\psi _l^\varepsilon }}
 \mathrel{\left | {\vphantom {{\psi _l^\varepsilon  } {\psi _l^\varepsilon  }}}
 \right. \kern-\nulldelimiterspace}
 {{\psi _l^\varepsilon  }} \right\rangle _\varepsilon $, is
\begin{equation}\label{3}
\begin{array}{l}
 \frac{{R_\varepsilon\left( {\lambda _{i,l}^\varepsilon  } \right)}}{{\left\langle {{\psi _l^\varepsilon  }}
 \mathrel{\left | {\vphantom {{\psi _l^\varepsilon  } {\psi _l^\varepsilon  }}}
 \right. \kern-\nulldelimiterspace}
 {{\psi _l^\varepsilon  }} \right\rangle_\varepsilon }} = \tilde R_\varepsilon\left( {\lambda _{i,l}^\varepsilon  } \right)  \\
 \cong -\lambda _{o,l}^\varepsilon   + \sum\limits_{i = 1}^M {\lambda
_{i,l}^\varepsilon  \left\{ {\frac{{\left\langle {\psi
_l^\varepsilon  } \right|\Theta \left( x \right)\left| {\psi
_l^\varepsilon  } \right\rangle_\varepsilon }}{{\left\langle {{\psi
_l^\varepsilon  }}
 \mathrel{\left | {\vphantom {{\psi _l^\varepsilon  } {\psi _l^\varepsilon  }}}
 \right. \kern-\nulldelimiterspace}
 {{\psi _l^\varepsilon  }} \right\rangle }_\varepsilon} + d_i^\varepsilon  } \right\}}  \\
 \end{array}
\end{equation}

Here, (7) is $ \tilde R_\varepsilon\left( {\lambda
_{i,l}^\varepsilon } \right) \cong \tilde I_\varepsilon^{FCC} \left(
{\lambda _{i,l}^\varepsilon ,d_i^\varepsilon } \right) -
\Im_\varepsilon \left( {\lambda _{o,l}^\varepsilon ,\lambda
_{i,l}^\varepsilon  } \right);i = 1,...,M $, where the re-scaled
target FCC and the \textit{potential energy} are $ \tilde
I_\varepsilon ^{FCC} \left( {\lambda _{i,l}^\varepsilon
,d_i^\varepsilon  } \right) \ $, and, $ \ \Im_\varepsilon \left(
{\lambda _{o,l}^\varepsilon ,\lambda _{i,l}^\varepsilon } \right) $,
respectively. A steepest descent procedure $  \frac{{\partial \tilde
R_\varepsilon\left( {\lambda _{i,l}^\varepsilon  }
\right)}}{{\partial \lambda _{i,l}^\varepsilon  }} \to 0 $ along the
gradient of the LM's yields ``optimal" values of the LM's
\begin{equation}\label{3}
\frac{{\partial \left[ {\tilde I_\varepsilon^{FCC} \left( {\lambda _{i,l}^\varepsilon  ,d_i^\varepsilon  } \right) - \Im_\varepsilon \left( {\lambda _{o,l}^\varepsilon  ,\lambda _{i,l}^\varepsilon  } \right)} \right]}}{{\partial \lambda _{i,l}^\varepsilon  }} \to 0;i = 1,...,M \\
\end{equation}
The optimization procedure is carried out till the target values are
achieved.  The steepest descent procedure requires the analytical
values of $ \frac{{\partial \tilde R_\varepsilon\left( {\lambda
_{i,l}^\varepsilon  } \right)}}{{\partial \lambda _{i,l}^\varepsilon
}} $, and thus, $ \frac{{\partial \lambda _{o,l}^\varepsilon
}}{{\partial \lambda _{i,l}^\varepsilon }};i = 1,...,M $. Left
multiplying (3) for the \textit{energy state} $ \varepsilon $ and
iteration level $ l $ by $ {\psi _l^\varepsilon } $ and integrating,
yields $ \frac{{\partial \lambda _{o,l}^\varepsilon }}{{\partial
\lambda _{i,l}^\varepsilon }} = \frac{{d_{i,l}^\varepsilon
}}{{\left\langle {{\psi _l^\varepsilon }}
 \mathrel{\left | {\vphantom {{\psi _l^\varepsilon  } {\psi _l^\varepsilon  }}}
 \right. \kern-\nulldelimiterspace}
 {{\psi _l^\varepsilon  }} \right\rangle_\varepsilon }};i = 1,...,M $.  The theory of the semi-supervised learning paradigm based on the \textit{Fisher game} is summarized
 by
the pseudo-code in Algorithm 1.
\subsection{Physical interpretations}
The \textit{Fisher game} constitutes a self-consistent information
theoretic optimization procedure, with a quantum mechanical
connotation.  The above theory contains three interesting
observations.  First, the commencement of each \textit{iteration
loop} corresponds to the juncture at which the observer initiates
measurements.  Next, as the iterative process advances, the FCC
approaches a minimum.  This corresponds to an increase in the
uncertainty at the location of the observer.

Finally, at the termination of the $ l^{th} $ iteration loop, the
condition (8) that yields the "optimal" LM's is the statement of a
\textit{contract} between the \textit{demon} and the observer,
whereby, the \textit{demon} makes the last move in the iterative
Fisher game. This implies that the \textit{participatory observer}
acquires a state of maximum uncertainty (minimum Fisher
information).  Such a \textit{contract} is the underlying basis for
determining the "optimal" LM's, corresponding to amplitudes that
decrease the FCC at the termination of each iteration level.

Scenarios of such type cannot be modeled within the framework of
traditional game theory [17], thus, justifying the use of the
\textit{bounded rationality theory} to study the game theoretic
aspects of the \textit{Fisher game}.  A future publication studies
the information landscape and its relation to the \textit{Fisher
game}.

\begin{algorithm}
\caption{Inverse Problem of Statistics}
\begin{algorithmic}
\STATE \textbf{PROCEDURE FOR EACH ENERGY STATE $ \varepsilon $}

\STATE \textbf{INITIALIZATION}
\\
1.Solve TISE (1) for known physical potential $ V(x) $ as an
eigenvalue problem in the event space $ \left[ {a,b} \right] $.
Obtain incomplete constraints $ d_i^\varepsilon;i = 1,...,M $.
\\
2.  Solve TIS-lE (3) for arbitrary Lagrange multipliers $
\lambda_{i,arbitrary}^\varepsilon; i=1,...,M $. Obtain amplitudes $
\psi_{i,l=1}^\varepsilon $.
\\
3.  Input tolerance parameter $ \delta_i;i=1,...,M $

\STATE \textbf{ALGORITHM FOR $ l^{th}\geq 1 $ ITERATION LOOP}
\\
1. Obtain $ \psi _l^\varepsilon $ by
 solving the TIS-lE (3) as an eigenvalue problem, using the optimized LM's $
 \lambda_{i,l-1}^\varepsilon $ obtained from the $ (l-1)^{th} $
 iteration loop.
 \\
 2.  Minimize the re-scaled \textit{residue} $ \tilde R_\varepsilon\left( {\lambda _{i,l}^\varepsilon  } \right)
 $ (7), to obtain optimized LM's that correspond to $ \psi _l^\varepsilon $

$ \lambda_{i,l}^\varepsilon \leftarrow  \frac{{\partial \left[
{\tilde I_\varepsilon^{FCC} \left( {\lambda _{i,l}^\varepsilon
,d_i^\varepsilon  } \right) - \Im_\varepsilon \left( {\lambda
_{o,l}^\varepsilon  ,\lambda _{i,l}^\varepsilon  } \right)}
\right]}}{{\partial \lambda _{i,l}^\varepsilon  }} $
\\
3. Solve TIS-lE (3) as an eigenvalue problem with LM's $
\lambda_{i,l}^\varepsilon $.  Obtain moments $ d_{i,l}^\varepsilon;i
= 1,...,M $.
\\

\STATE \textbf{IF}
\\
$ \left| {d_i^\varepsilon   - d_{i,l}^\varepsilon  } \right| >
\delta_i;i=1,...,M $
\\
$l=l+1$ \STATE \textbf{RETURN}
\\
\STATE \textbf{ELSE IF}
\\
 $
\left| {d_i^\varepsilon   - d_{i,l}^\varepsilon  } \right| \le
\delta_i;i=1,...,M $
\\
\STATE \textbf{END IF}

\end{algorithmic}
\end{algorithm}

\subsection{Numerical results}

The asymmetric harmonic oscillator (AHO) potential $ V\left( x
\right) = x + \frac{{x^2 }}{2} $ is chosen as TISE physical
potential. The TISE with the AHO potential is solved as an
eigenvalue problem for 201 data points within the event space $
\left[ {a=-1,b=1} \right] $ for different \textit{energy states}.
Boundary conditions on the amplitudes, $ \psi ^\varepsilon  \left( a
\right) = \psi ^\varepsilon  \left( b \right) = 0 $, are enforced.
An empirical pseudo-potential of the form $ V_\varepsilon \left( x
\right) = \tilde \lambda _1^\varepsilon x + \tilde \lambda
_2^\varepsilon $, that approximates the TISE AHO physical potential
is specified. Here, $ \tilde \lambda _i^\varepsilon = \frac{{\lambda
_i^\varepsilon }}{4} $. In this case, $ M=2, N=201 $.

The values of the incomplete constraints are $ d_1^{\varepsilon =
0,1} = ({\rm  - 0}{\rm .0344},{\rm 0}{\rm .0097,}), and,
d_2^{\varepsilon = 0,1}  = ({\rm 0}{\rm .1302},{\rm 0}{\rm .2809})
$. The final values of the LM's are $ \tilde \lambda _1^{\varepsilon
= 0,1} = ({\rm 1}{\rm .0892},{\rm 1}{\rm .1121}) $, and, $ \tilde
\lambda _2^{\varepsilon  = 0,1}  = ({\rm 0}{\rm .6635},{\rm 0}{\rm
.7175})\ $, respectively.  The inferred total energy eigenvalues are
$ \ E_{\inf }^{\varepsilon  = 0,1}  = ({\rm 2}{\rm .5333},{\rm
10}{\rm .0769}) \ $. The corresponding TISE total energy eigenvalues
are $ \ E_{exact}^{\varepsilon  = 0,1} = (2.5152,{\rm 10}{\rm
.0147}) \ $. Fig. 1 depicts the inferred \textit{pdf}'s overlaid
upon the TISE solution.  Here, the Maxwellian distribution
corresponds to $ \varepsilon =0 $, and, double peaked \textit{pdf}
corresponds to the first \textit{excited state} $ \varepsilon =1 $.
Note that the inferred \textit{pdf}'s almost exactly coincide with
the TISE solutions.
\begin{figure}[thpb]
\includegraphics[scale=0.55]{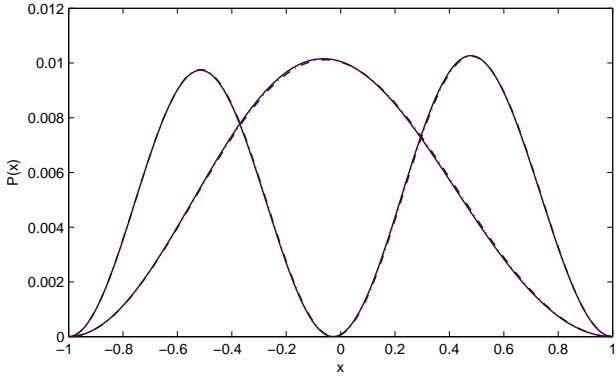}

\caption{Inferred TIS-lE \textit{pdf} (dash-dots) and exact TISE
\textit{pdf} (solid lines)}
\end{figure}

\section{Projection Strategy}
Consider $ \ M \ $ constraints  $ \ d_1^\varepsilon
,...,d_M^\varepsilon \ $. In a discrete setting, these are
expectation values of a random variable $ \ x_{i,n} ;n = 1,...,N \
$:
\begin{equation}\label{3}
\ d_i^\varepsilon  = \sum\limits_{n = 1}^N {p_n^\varepsilon x_{i,n}
} ;i = 1,...,M.\
\end{equation}
The \textit{pdf} $ \left| p^\varepsilon \right\rangle \in \Re ^N $
is a  \textit{ket}, where $ \left| n \right\rangle ;n = 1,...,N $ is
the standard basis in $ \Re ^N $, is expressed as $ \left|
p^\varepsilon \right\rangle  = \sum\limits_{n = 1}^N {\left| n
\right\rangle \left\langle {n}
 \mathrel{\left | {\vphantom {n p}}
 \right. \kern-\nulldelimiterspace}
 {p^\varepsilon} \right\rangle }  = \sum\limits_{n = 1}^N {p_n^\varepsilon \left| n \right\rangle }
 $.
The \textit{ket} of observable's is expressed as $ \left|
d^\varepsilon \right\rangle \in \Re ^{M + 1} $ with components $
d_1^\varepsilon ,...,d_M^\varepsilon,1 $, and, an operator $ A:\Re
^N  \to \Re ^{M + 1} $ given by $ A = \sum\limits_{n = 1}^N {\left|
{x_n } \right\rangle \left\langle n \right|} \ $. Defining vectors $
\left| {x_n } \right\rangle \in \Re ^{M + 1} ;n = 1,...,N $ as the
expansion $ \left| {x_n } \right\rangle  = \sum\limits_{i = 1}^{M +
1} {\left| i \right\rangle } \left\langle {i}
 \mathrel{\left | {\vphantom {i {x_n }}}
 \right. \kern-\nulldelimiterspace}
 {{x_n }} \right\rangle  = \sum\limits_{i = 1}^{M + 1} {x_{i,n} } \left| i \right\rangle
 $, where $ \ i \ $ is a basis vector in $  \Re ^{M + 1}  $, (9)
acquires the compact form
\begin{equation}\label{7} \left| {d^\varepsilon  } \right\rangle =
A\left| {p^\varepsilon } \right\rangle ;A:\Re ^N  \to \Re ^{M + 1}
.\
\end{equation}

The physical significance of the constraint operator $ A $ in (10)
is as follows. Inference of the \textit{pdf} and the TIS-lE
pseudo-potential in (3) from physical observables is achieved by
specifying $ V\left( x \right) = \sum\limits_{i = 1}^M {\lambda _i }
x^i $.  In a discrete setting, $ x_n^i  \to x_{i,n} ;i = 1,...,M;n =
1,...,N $. The $ x_{i,n} $ constitute the elements of the $ M $ rows
and $ N $ columns of the operator $ A $, and represent the spatial
elements of the TIS-lE pseudo-potential in matrix form. The unity
element in $ \left| d^\varepsilon \right\rangle \in \Re ^{M + 1} $
enforces the normalization constraint of the probability density $
\left| {p^\varepsilon  } \right\rangle $.

\textit{The operator $ A $ is independent of the host pdf, and thus,
the \textit energy state}. This may be mitigated by defining
\begin{equation}\label{3}
\tilde A^\varepsilon = A + k^\varepsilon\left| {d^\varepsilon }
\right\rangle \left\langle I \right| \
\end{equation}
Here,  $ \left\langle I \right| $ is a $ 1 \times N  $ \textit{bra},
and, $ k^\varepsilon \neq -1 $ is a constant parameter introduced to
adjust the condition number of $ \tilde A^\varepsilon $, and hence
its sensitivity to perturbations.  In (11), dependence upon the host
\textit{pdf} is "injected" into the operator $  \tilde A^\varepsilon
$ by the incorporation of $ \left| {d^\varepsilon } \right\rangle $.
Specifically, each element of the \textit{ket} $ \left|
{d^\varepsilon } \right\rangle $ is defined by $ \sum\limits_{n =
1}^N {p_n^\varepsilon  } x_n^i ;i = 1,...,M $.

Thus, (10) becomes $ \tilde A^\varepsilon  \left| {p^\varepsilon  }
\right\rangle  = \left| {d^\varepsilon  } \right\rangle  +
\left\langle {k^\varepsilon\left| {d^\varepsilon  } \right\rangle
\left\langle I \right|} \right\rangle _\varepsilon $.  Expanding $
\left\langle {k\left| {d^\varepsilon } \right\rangle \left\langle I
\right|} \right\rangle _\varepsilon = k\left| {d^\varepsilon  }
\right\rangle \left\langle I \right|\left. {p^\varepsilon  }
\right\rangle $, and evoking the \textit{pdf} normalization, $
\left\langle I \right|\left. {p^\varepsilon  } \right\rangle  = 1 $,
yields
\begin{equation}\label{4}
\left| {\tilde d^\varepsilon  } \right\rangle  = \left(
{k^\varepsilon + 1} \right)\left| {d^\varepsilon  } \right\rangle  =
\tilde A^\varepsilon  \left| {p^\varepsilon  } \right\rangle ;\tilde
A^\varepsilon  :\Re ^N  \to \Re ^{M + 1} .\
\end{equation}

The operator $ \tilde A^\varepsilon \ $is ill-conditioned and
rectangular.  Thus, (12) becomes:
\begin{equation}\label{3}
\left| {p^\varepsilon  } \right\rangle  = \left( {\tilde
A^\varepsilon  } \right)^{ - 1} \left| {\tilde d^\varepsilon  }
\right\rangle  + \left| {p^{\varepsilon '} } \right\rangle, \
\end{equation}
where, $ {\left( {\tilde A^\varepsilon  } \right)^{ - 1} } $ is the
pseudo-inverse [18] of $ \tilde A^\varepsilon \ $, and lies in $
range\left( {\tilde A^\varepsilon } \right) $. All necessary data
dependent information resides in $ \left( {\tilde A^\varepsilon }
\right)^{ - 1} \left| {\tilde d^\varepsilon } \right\rangle $.

The \textit{null space} term in (13) is of particular importance
since the code is embedded into it via unitary projections.  Here, $
\left| {p^{\varepsilon '} } \right\rangle  \in null\left( {\tilde
A^\varepsilon  } \right) $ is explicitly data independent. However,
it is critically dependent on the solution methodology employed to
solve (13).  The operator $ G^\varepsilon = \tilde A^{\varepsilon
\dag} \tilde A^\varepsilon  $ is introduced. Here, $ \tilde
A^{\varepsilon \dag} $ is the conjugate transpose of $ \tilde
A^\varepsilon  $. Projection of the covert information into $
null\left( {G^\varepsilon  } \right) $ instead of $ null\left(
{\tilde A^\varepsilon } \right) $, leads to increased instability of
the eigenstructure,  which is exploited to increase the security of
the covert information [3, 9].

Given the operator $ \tilde A^\varepsilon $ and the probability
vector $ \left| p^\varepsilon \right\rangle $, whose inference is
described in Section 2, the normalized eigenvectors corresponding to
the eigenvalues in the \textit{null space} of $ G^\varepsilon $
having value zero (\textit{zero eigenvalues}) are defined as  $
\left| {\eta _n^\varepsilon  } \right\rangle ;n = 1,...,N - (M + 1)
$. Here, $ \left| {\eta _n^\varepsilon  } \right\rangle $, defined
as the basis of $ null\left( {G^\varepsilon  } \right) $, are
evaluated using SVD [18]. To introduce cryptographic keys
(cryptographic primitives), an operator $ \tilde G^\varepsilon $ is
formed by perturbing select elements of $ \ G^\varepsilon \ $ by $
\delta G_{i,j}^\varepsilon $. Here, $ \delta G_{i,j}^\varepsilon $
is a perturbation to the element inhabiting the $ i^{th} $ row and $
j^{th} $ column of the operator $ G^\varepsilon $.  In
\textit{symmetric} cryptography, only a single element of $
G^\varepsilon \ $ is perturbed. The security of the code may be
ensured by adopting an \textit{asymmetric} cryptographic strategy.
Here, more than one element of $ G^\varepsilon $ is perturbed.

The extreme sensitivity to perturbations of $ G^\varepsilon \ $
causes the eigenstructure of $ {\tilde G^\varepsilon = \
G^\varepsilon + \delta G_{i,j}^\varepsilon } \ $ to substantially
differ from that of $ {G^\varepsilon } $, even for infinitesimal
perturbations. The values of $ \left| {\tilde \eta _n^\varepsilon  }
\right\rangle ;n = 1,...,N - (M + 1) $, the basis of $ null(\tilde
G^\varepsilon) $, are evaluated using SVD.  The unitary operators of
decryption $ U_{dec}^\varepsilon \ $ (without perturbations) and $
\tilde U_{dec}^\varepsilon \ $ (with perturbations), $
U_{dec}^\varepsilon, \tilde U_{dec}^\varepsilon   :\Re ^N \to \Re
^{N - (M + 1)} $, and, the corresponding encryption operators $
U_{enc}^\varepsilon, \tilde U_{enc}^\varepsilon \ :\Re ^{N - (M +
1)} \to \Re ^N $ for the \textit{energy state} $ \varepsilon $ are
\begin{equation}\label{7}
\begin{array}{l}
 U_{dec}^\varepsilon   = \sum\limits_{n = 1}^{N - M - 1} {\left| n \right\rangle \left\langle {\eta _n^\varepsilon  } \right|} ; \\
 \tilde U_{dec}^\varepsilon   = \sum\limits_{n = 1}^{N - M - 1} {\left| n \right\rangle \left\langle {\tilde \eta _n^\varepsilon  } \right|} , \\
 and, \\
 U_{enc}^\varepsilon   = U_{dec}^{\varepsilon \dag }  = \sum\limits_{n = 1}^{N - M - 1} {\left| {\eta _n^\varepsilon  } \right\rangle \left\langle n \right|} ;\\
 \tilde U_{enc}^\varepsilon   = \tilde U_{dec}^{\varepsilon \dag }= \sum\limits_{n = 1}^{N - M - 1} {\left| {\tilde \eta _n^\varepsilon  } \right\rangle \left\langle n \right|} ,  \\
 \end{array}
\end{equation}
respectively.
\subsection{Encryption}
Given a code $ \left| q^\varepsilon  \right\rangle \in \Re ^{N - (M
+ 1)} $ to be encrypted in the energy state $ \varepsilon $ , the $
N - \left( {M + 1} \right) \ $ components are given by $
\left\langle {n}
 \mathrel{\left | {\vphantom {n {q^\varepsilon  }}}
 \right. \kern-\nulldelimiterspace}
 {{q^\varepsilon  }} \right\rangle  = q_n^\varepsilon  ;n = 1,...,N - \left( {M + 1} \right)
$.  The \textit{pdf} of the embedded code is:
\begin{equation}\label{3}
\left| {p_c^\varepsilon  } \right\rangle  = \tilde
U_{enc}^\varepsilon \left| {q^\varepsilon  } \right\rangle  =
\sum\limits_{n = 1}^{N - M - 1} {\left| {\tilde \eta _n^\varepsilon
} \right\rangle \left\langle {n}
 \mathrel{\left | {\vphantom {n {q^\varepsilon  }}}
 \right. \kern-\nulldelimiterspace}
 {{q^\varepsilon  }} \right\rangle }.
\end{equation}

The total \textit{pdf} comprising the host \textit{pdf} and the
\textit{pdf} of the code is
\begin{equation}\label{7}
\left| {\tilde p^\varepsilon  } \right\rangle  = \left|
{p^\varepsilon  } \right\rangle  + \left| {p_c^\varepsilon  }
\right\rangle. \
\end{equation}
Note that since $ \left| {p_c^\varepsilon  } \right\rangle  \in
null\left( {\tilde A^\varepsilon  } \right) \ $, $ \ \tilde
A^\varepsilon \left| {p_c^\varepsilon } \right\rangle = 0 \ $.

\subsection{Transmission}
Information may be transferred from the encrypter to the decrypter
in two separate manners , via a \textit{public channel}.  The first
mode is to transmit the constraint operators $ \tilde A^\varepsilon
$ and the total \textit{pdf}'s $  \left| {\tilde p^\varepsilon  }
\right\rangle $. An alternate mode is to transmit the LM's obtained
on solving the \textit{Fisher game}(Section 2), and, the total
\textit{pdf}'s $ \left| {\tilde p^\varepsilon  } \right\rangle $.
Owing to the large dimensions of the constraint operators $ \tilde
A^\varepsilon $, the latter transmission strategy is more
attractive.

The values of parameters $ k^\varepsilon $ for each \textit{energy
state}, and, the cryptography key/keys are transmitted through a
\textit{secure/covert channel}. The cryptographic primitives are
labeled in order to identify the elements of the operator $ \
G^\varepsilon $ that are perturbed. In the case of
\textit{asymmetric} cryptography, some of the keys may be declared
public, while keeping the remainder private [4].
\textit{Asymmetric} cryptography provides greatly enhanced security
to the covert information, and, provides protection against attacks,
such as \textit{plaintext attacks} [3, 4, 9].

\subsection{Decryption}
The  decrypter and encrypter have an \textit{a-priori} "agreement"
concerning the nature of TIS-lE pseudo-potential, and, the number of
\textit{energy states}.  The legitimate receiver recovers the
key/keys $ \delta G_{i,j}^\varepsilon $ and the parameter $
k^\varepsilon $ from the \textit{covert channel}.  The operators $
\tilde A^\varepsilon  \ $, $ \ G^\varepsilon \ $, and, $ \tilde
G^\varepsilon  \ $ are constructed. The host \textit{pdf} may be
recovered in two distinct manners, depending upon the transmission
strategy employed. Note that both methods to reconstruct the host
\textit{pdf} require the total \textit{pdf} $ {\tilde p^\varepsilon
} $ to be provided by the encrypter. First, the scaled incomplete
constraints, defined in (12), are obtained by solving $ \
\left\langle i \right|\tilde A^\varepsilon \left| {\tilde
p^\varepsilon  } \right\rangle \ = \ \left| {\tilde d^\varepsilon }
\right\rangle \ $.  Here, $ \ i \ $ is a basis vector in $  \Re ^{M
+ 1} \ $.  This procedure is possible because $ \left|
{p_c^\varepsilon  } \right\rangle \in null\left( {\tilde
A^\varepsilon  } \right) $. Thus, $ \ \tilde A^\varepsilon \left|
{p_c^\varepsilon } \right\rangle = 0 \ $. The host \textit{pdf} are
then computed for each \textit{energy state} by solving the
\textit{Fisher game}, using the re-scaled set of incomplete
constraints. Alternatively, the host \textit{pdf} may be obtained by
solving the TIS-lE (3) as an eigenvalue problem, given the values of
the LM's $ \lambda _i^\varepsilon \ ; i=1,...,M $, and, the event
space (Section 2). Both methods allow the reconstructed host
\textit{pdf}'s  to be obtained with a high degree of precision.  The
code \textit{pdf} is recovered using
\begin{equation}\label{3}
\left| {p_{rc}^\varepsilon  } \right\rangle  = \left| {\tilde
{p^\varepsilon }} \right\rangle  - \left| {p^\varepsilon }
\right\rangle. \
\end{equation}
The encrypted code is recovered by the operation
\begin{equation}\label{3}
\left| q_r^\varepsilon \right\rangle  = \tilde U_{dec}^\varepsilon
\left| {p_{rc}^\varepsilon  } \right\rangle = \sum\limits_{n = 1}^{N
- M - 1} {\left| n \right\rangle \left\langle {{\tilde \eta
_n^\varepsilon }}
 \mathrel{\left | {\vphantom {{\tilde \eta _n } {p_{rc}^\varepsilon  }}} \right. \kern-\nulldelimiterspace}
 {{p_{rc}^\varepsilon }} \right\rangle } \
\end{equation}

The thresholds for the cryptographic keys is accomplished by the
designer, who performs a simultaneous encryption/decryption without
effecting perturbations to the operator $ G^\varepsilon $.  The host
\textit{pdf}'s are inferred from the \textit{Fisher game}. The code
$ \left| {q^\varepsilon } \right\rangle $ having dimension $ N-(M+1)
$ is formed.  The designer implements (15)-(18) for each
\textit{energy state} $ \varepsilon $.  The threshold for the
cryptographic key/keys is $  \delta ^\varepsilon = \left\| {\left|
{q^\varepsilon } \right\rangle - \left| {q_r^\varepsilon }
\right\rangle } \right\| $.   Hardware independence is demonstrated
by performing the encryption on an IBM RS-6000 workstation cluster,
and, decryption on an IBM Thinkpad running MATLAB v 7.01. The
encryption/decryption strategy is critically dependent upon the
exact compatibility of the routines to calculate the basis $ \left|
{\tilde \eta _n^\varepsilon } \right\rangle $ and the eigenvalue
solvers, available to the encrypter and decrypter.

\section{Numerical Examples}
The encryption/decryption strategy is tested using the
\textit{energy state} dependent model vis-\'{a}-vis an
\textit{energy state} independent model [9], for the case of
\textit{asymmetric} cryptography.  These are characterized by the
constraint operators $ \tilde A^\varepsilon $ (described in (11) and
(12)) for $ k^{\varepsilon=0,1}=-0.1 $, and, $ A $ (described in
(10)), respectively. The \textit{energy state} independent model
corresponds to the ground state Maxwellian distribution.

A random number generator generates code in $ \left[ {0,1} \right]
$.  Two identical \textit{kets} of the code having dimension $
N-(M+1)=198 $ are created for projection into the \textit{null
spaces} of the \textit{energy state} dependent operators $ \tilde
G^{\varepsilon=0,1} $, respectively. This "emulates" the selective
projection of a code comprising of a single \textit{ket }of
dimension $ 396 $, into the two \textit{energy state} of $ null(
\tilde G^\varepsilon $. For the \textit{energy state} independent
operator $ null(\tilde G) $, only a single \textit{ket} is
projected.  The cryptographic primitives are $ \delta G_{1,3} =
\delta \tilde G_{1,3}^{\varepsilon = 0,1} = 3.0e - 013 $ and $
\delta G_{2,7} = \delta \tilde G_{2,7}^{\varepsilon = 0,1}  = 7.0e -
013 $, respectively. All numerical examples in have a threshold for
perturbations $ \delta ^\varepsilon \sim 2.0e - 014 $. The condition
numbers, $ cond\left(  \bullet  \right) $ of the constraint
operators $ A $ and $ \tilde A^\varepsilon $ provides a measure of
the sensitivity to perturbations of the operators $ G= A^\dag  A $
and $ G^\varepsilon $.

Values of $ cond(A) $, $ cond(\tilde A^{\varepsilon=0}) $, and, $
cond(\tilde A^{\varepsilon=1}) $ are $ {\rm 3}{\rm .73048} $, $
3.{\rm 41742} $, and, $ {\rm 3}{\rm .37888} \ $, respectively. Going
by conventional logic, the \textit{energy state} independent model
should afford greater security to the covert information, owing to
the greater value of $ cond(A) $, vis-\'{a}-vis $ cond(\tilde
A^{\varepsilon=0,1}) $.  Numerical simulations reveal a dichotomy in
this regard.

A more relevant metric of the extreme sensitivity of $ null\left(
{G^\varepsilon } \right) $ to perturbations, induced by the
cryptographic keys $ \delta G_{i,j}^\varepsilon $, is the distortion
of the code \textit{pdf} $ \left| {p_{c,unpert}^\varepsilon }
\right\rangle $. Here, $ \left| {p_{c,unpert}^\varepsilon }
\right\rangle $ is evaluated from (15), using $ \eta_n^\varepsilon $
(the unperturbed basis of $ null\left( {G^\varepsilon } \right) $).
The distorted code \textit{pdf} is $ \left| {\tilde p_c^\varepsilon
} \right\rangle $, which is calculated from (15) using $ \tilde
\eta_n^\varepsilon $ (the perturbed basis of $ null\left( {\tilde
G^\varepsilon } \right) $), as described in Section 3.1.

For the \textit{energy state} dependent model, the \textit{RMS error
of encryption}is defined as: $ RMS_{enc}^{\varepsilon} =
\frac{{\left\| {err_{enc}^{\varepsilon} } \right\|}}{{\sqrt
{length\left( {err_{enc}^{\varepsilon} } \right)} }} $. Here, $
\left\| {\left| {err_{enc}^\varepsilon  } \right\rangle } \right\| \
= \left\| {\left( {\left| {p_c^{\varepsilon} } \right\rangle -
\left| {p_{c,unpert}^{\varepsilon} } \right\rangle } \right)}
\right\| \ $, and, $ length\left( {\left| {err_{enc}^\varepsilon  }
\right\rangle } \right) $ is the dimension of $ \left( {\left|
{p_c^\varepsilon  } \right\rangle  - \left|
{p_{c,unpert}^\varepsilon  } \right\rangle } \right) $.

A further quantitative metric of the degree of security of the
encrypted code is the RMS error of reconstruction between the
embedded code and the code reconstructed without the keys.  For the
\textit{energy state} dependent model, this is: $
RMS_{recon}^\varepsilon   = {\raise0.7ex\hbox{${\left\| {\left|
{err_{recon}^\varepsilon  } \right\rangle } \right\|}$}
\!\mathord{\left/
 {\vphantom {{\left\| {\left| {err_{recon}^\varepsilon  } \right\rangle } \right\|} {length\left( {\left| {err_{recon}^\varepsilon  } \right\rangle } \right)}}}\right.\kern-\nulldelimiterspace}
\!\lower0.7ex\hbox{${length\left( {\left| {err_{recon}^\varepsilon
} \right\rangle } \right)}$}};\left\| {\left|
{err_{recon}^\varepsilon  } \right\rangle } \right\| = \left\|
{\left| q \right\rangle  - \left| {q_{r1} } \right\rangle } \right\|
\ $.  Here, $ RMS_{recon}^{\varepsilon} $  provides a measure of the
error of recovery of the code by
 an \textit{unauthorized eavesdropper} who does not possess the keys
 $  \delta G_{i,j}^\varepsilon $, but, possesses the total \textit{pdf} and the code \textit{pdf}.  Such attackers are known as \textit{semi-honest adversaries}, since no attempt is made to distort the information transmitted via the
public channel.  In this case, the reconstructed code becomes $
\left| q_{r1}^\varepsilon \right\rangle  = \sum\limits_{n = 1}^{N -
M - 1} {\left| n \right\rangle \left\langle {{\eta_n^\varepsilon }}
 \mathrel{\left | {\vphantom {{\tilde \eta _n } {p_{rc}^\varepsilon  }}}
 \right. \kern-\nulldelimiterspace}
 {{p_{rc}^\varepsilon }} \right\rangle } \ $.

For the \textit{energy state} independent model, the values of $
RMS_{enc}=0.79869 $ and $ RMS_{recon}=0.81302 $.  The corresponding
values for the \textit{energy state} dependent model are $
RMS_{enc}^{\varepsilon=0,1}=(0.83878, 0.89651) $, and, $
RMS_{recon}^{\varepsilon=0,1}= {\rm 0}{\rm .84596},{\rm 0}{\rm
.88931} $, respectively.   The higher values of the $
RMS_{enc}^\varepsilon $ for the \textit{energy state} dependent
model explains the vastly enhanced degree of security it provides by
demonstrating a greater value of $ RMS_{recon}^{\varepsilon} $,
despite the value of $ cond(\tilde A^\varepsilon) $ being less than
$ cond(A) $. Simulations results for select values for the case of
the \textit{energy state} independent and dependent models are
described in Table 1 and Table 2, respectively. The reconstructed
code \textit{with} the keys is exactly similar to the original code.
\textit{On the other hand, the code reconstructed without the keys
bears no resemblance to the original code}.  The highly oscillatory
nature of the total \textit{pdf} (16) for the \textit{energy state}
dependent model depicted in Fig. 2, demonstrates the extreme
instability of the statistical coding process.

\section{Ongoing Work}
The \textit{Fisher game} has been extended to multi-dimensional and
temporal cases.   The model presented herein is in the process of being
amalgamated with existing quantum key distribution protocols [19],
to yield a hybrid statistical/quantum mechanical cryptosystem.  Such
a hybrid cryptosystem mitigates the current limitations of quantum
channels to transmit large amounts of data.  A \textit{covert} quantum key distribution protocol may be utilized for the secure delivery of the cryptographic primitives (the $
\delta G_{i,j}^\varepsilon $).   Finally, the statistical
encryption/decryption strategy has been modified to perform
privacy protection in statistical databases. These results will be
published elsewhere.
\begin{table}[h]
\caption{Energy State Independent Model}
\label{sample-table}
\begin{center}
\begin{tabular}{ll}
\multicolumn{1}{c}{\bf $ \left| {q }\right\rangle = \left| {q_{r}  } \right\rangle \ $} &\multicolumn{1}{c}{\bf $ \left| {q_{r1}  } \right\rangle $ } \\
\hline \\

 0.23813682639005 &    -0.00668168344388 \\
  0.69913526160795 &   0.20008072567388 \\
  0.27379424177629 &   -0.14186802540956 \\
  0.90226539453884 &  0.36853370671177 \\
\end{tabular}
\end{center}
\end{table}

\begin{table}[h]
\caption{Energy State Dependent Model}
\label{sample-table}
\begin{center}
\begin{tabular}{ll}
\multicolumn{1}{c}{\bf $ \left| {q^{\varepsilon} }\right\rangle = \left| {q_{r}^{\varepsilon}  } \right\rangle \ $} &\multicolumn{1}{c}{\bf $ \left| {q_{r1}^\varepsilon  } \right\rangle $ } \\
\hline \\

$ \varepsilon=0 $ \textit{Zero-energy/ground state} \\

 0.23813682639005 &    -0.26776249759842 \\
  0.69913526160795 &   0.77862610842042 \\
  0.27379424177629 &    -1.16636783859136 \\
  0.90226539453884 &  0.02881517541356 \\
\\

$ \varepsilon=1 $ \textit{First excited state} \\

0.23813682639005 &    1.25161826270042 \\
  0.69913526160795 &   -3.255410114151938e-005 \\
  0.27379424177629 &    -0.11041660156776 \\
  0.90226539453884 &  0.61665920565232 \\
\end{tabular}
\end{center}
\end{table}

\begin{figure}[thpb]
\centering
\includegraphics[scale=0.50]{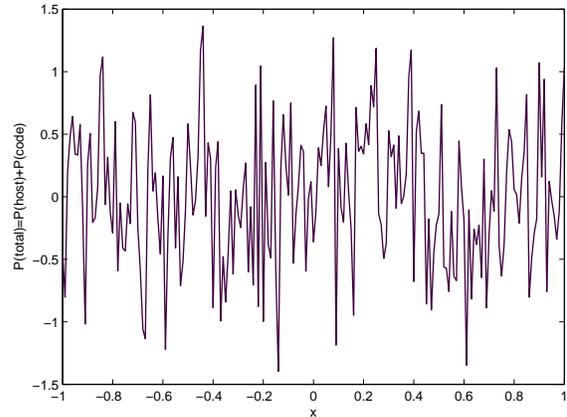}
\caption{"Chaotic" Nature of the Total Pdf, Eq. (16)}
\end{figure}


\subsubsection*{Acknowledgements}

This work was supported by \textit{RAND-MSR contract} \textit{CSM-DI
$ \ \& $ S-QIT-101107-2005}.  Gratitude is expressed to B. R.
Frieden, A. Plastino, and, B. Soffer for helpful discussions.


\nocite{ex1,ex2}
\bibliographystyle{latex8}
\bibliography{latex8}

\end{document}